\documentclass{PoS}

\usepackage{color, colortbl}
\definecolor{LightBlue}{rgb}{0.9,0.9,1}

\title{Lattice NRQCD study of quarkonium at non-zero temperature}

\ShortTitle{NRQCD study of quarkonium at $T \neq 0$}

\author{\speaker{Seyong Kim} \\
        Department of Physics, Sejong University, Seoul 143-747, Korea \\
        E-mail: \email{skim@sejong.ac.k}}

\author{Peter Petreczky \\
        Physics Department, Brookhaven National Laboratory, Upton,
        NY11973, USA \\
        E-mail: \email{petreczk@quark.phy.bnl.gov}}

\author{Alexander Rothkopf \\
        Institute for Theoretical Physics, Heidelberg University,
        Philosophenweg 16, 69120 Heidelberg, Germany \\ 
        E-mail: \email{rothkopf@thphys.uni-heidelberg.de}}

\abstract{To study the in-medium modification of quarkonium properties,
charmonium correlators at $140.4 (\beta =6.664) \le T \le 221 (\beta =
7.280)$ (MeV) are calculated using the NRQCD formalism on $48^3 \times
12$ gauge configurations with dynamical $N_f = 2 + 1$ flavors of
Highly Improved Staggered Quarks (HISQ). To determine the "zero energy
shift" for these lattices, we perform a fine zero temperature scan
($\beta =6.664, 6.740, 6.800, 6.880, 6.950, 7.030, 7.150$ and
$7.280$). We find that the temperature
dependence of charmonium correlators is stronger than the temperature
dependence of bottomonium correlators in a given channel. This fits
into the expected pattern of sequential quarkonium melting.}

\FullConference{The 33rd International Symposium on Lattice Field Theory\\
		14 -18 July 2015\\
		Kobe International Conference Center, Kobe, Japan*}

\begin{document}

\section{Introduction}

Quarkonium is expected to play an important role as an indicator for
quark-gluon plasma (QGP) formation \cite{Matsui:1986dk}. A
quantitative understanding of how quarkonium behaves around the
transition temperature based on first principle calculations is a long
sought goal of many studies. Recently, bottomonium correlators were
studied within a lattice formulation of non-relativistic QCD (NRQCD)
in the background of gauge fields simulated in a non-zero temperature
environment with dynamical quarks. Spectral functions of various
channels have been reconstructed with Bayesian methods and were
scrutinized for in-medium modification
\cite{Aarts:2010ek,Aarts:2011sm,Aarts:2012ka,Aarts:2013kaa,Aarts:2014cda,Kim:2014nda,Kim:2014iga}.
There are several advantages of this approach: (1) NRQCD is an
effective field theory of QCD in which momenta above the soft scale
$mv$ including those of the order of the heavy quark rest mass are
``integrated out''.  It represents a systematic series expansion in
powers of $v$ (heavy quark velocity in the rest frame of quarkonium)
\cite{Bodwin:1994jh}. The difficulty associated with the lattice
conditions for a heavy quark that $a$ (lattice spacing) $\ll
\frac{1}{M} \ll La $ (lattice extent) is turned into an advantage as
basis of an effective description. (2) The kernel in the spectral
representation of NRQCD quarkonium correlators is temperature
independent and the problematic constant contribution to the spectral
function in relativistic QCD
\cite{Umeda:2007hy,Aarts:2002cc,Petreczky:2008px} is absent (note that
quarkonium state itself is not in thermal equilibrium). Note also that
NRQCD calculations of heavy quark propagators constitute an initial
value problem, not a boundary value problem like in the case of a
relativistic calculation. This reduces the computational effort
significantly.

The studies in
Refs.~\cite{Aarts:2010ek,Aarts:2011sm,Aarts:2012ka,Aarts:2013kaa,Aarts:2014cda}
use anisotropic lattices with a fixed lattice spacing (at a somewhat
heavy pion mass of $\sim 400$ MeV) and the system temperature is
increased by decreasing the number of Euclidean time slices. They
employ the Maximum Entropy Method (MEM) \cite{Asakawa:2000tr} for
spectral function reconstructions. The work in Ref.~\cite{Kim:2014iga}
uses isotropic lattices with variable lattice spacing (at nearly
physical pion and kaon masses). The system temperature is increased by
decreasing the lattice spacing. In Ref. \cite{Kim:2014iga} both MEM
and a new Bayesian Reconstruction (BR) method \cite{Burnier:2013nla}
are used for spectral function reconstruction.

These two studies use different lattice setups and employ different
spectral reconstruction methods. A comparison of such different
approaches may give us better understanding on the reconstruction
uncertainties of spectral functions via Bayesian methods in general.
In both \cite{Aarts:2011sm,Aarts:2014cda} and \cite{Kim:2014iga} a
consistent ``sequential suppression'' pattern of $\Upsilon (1S),
\Upsilon (2S)$ and $\Upsilon (3S)$ state has been observed in spectral
functions of the $\Upsilon$ channel, which were reconstructed both
with MEM and the BR method. Thus lattice NRQCD studies of S-wave
bottominum at $T \neq 0$ are on firm grounds. On the other hand for
P-wave bottomonium, the spectral functions reconstructed with MEM show
a ``melting'' behavior immediately close to $T_c$
\cite{Aarts:2013kaa,Aarts:2014cda,Kim:2014iga} while those
reconstructed with the BR method show survival of peaked structures
for the ground state up to $1.61 T_c$.

\begin{table}
\centering
\label{tab:latticedetail}
\begin{tabular}{llccccc}
\hline\hline
        \multicolumn{1}{c}{$M_ba$} &
        \multicolumn{1}{c}{$M_ca$} &
        \multicolumn{1}{c}{$\beta$} &
        \multicolumn{1}{c}{T(MeV)} &
        \multicolumn{1}{c}{T/$T_c$} &
	\multicolumn{1}{c}{$a^{-1}_\tau$ (fm)} &
        \multicolumn{1}{c}{No. of Conf.(analyzed)} \\
\hline
\rowcolor{LightBlue} 2.759 & 0.7566 &{ 6.664} & 140.40 & 0.911 & 0.1169 & 400 \\     
2.667 & 0.7314 & 6.700 & 145.32 & 0.944 & 0.1130 & 400 \\
\rowcolor{LightBlue}2.566 & 0.7035 &{ 6.740} & 150.97 & 0.980 & 0.1087 & 400 \\
2.495 & 0.6841 & 6.770 & 155.33 & 1.008 & 0.1057 & 400 \\
\rowcolor{LightBlue}2.424 & 0.6657 &{ 6.800} & 159.33 & 1.038 & 0.1027 & 400 \\
2.335 & 0.6403 & 6.840 & 165.95 & 1.078 &0.09893 & 400 \\
\rowcolor{LightBlue}2.249 & 0.6167 &{ 6.880} & 172.30 & 1.119 &0.09528 & 400 \\
2.187 & 0.5996 & 6.910 & 177.21 & 1.151 &0.09264 & 400 \\
\rowcolor{LightBlue}2.107 & 0.5776 &{ 6.950} & 183.94 & 1.194 &0.08925 & 400 \\
2.030 & 0.5566 & 6.990 & 190.89 & 1.240 &0.08600 & 400 \\
\rowcolor{LightBlue}1.956 & 0.5364 &{ 7.030} & 198.08 & 1.286 &0.08288 & 400 \\
1.835 & 0.5030 & 7.100 & 211.23 & 1.371 &0.07772 & 400 \\
\rowcolor{LightBlue}1.753 & 0.4806 &{ 7.150} & 221.08 & 1.436 &0.07426 & 400 \\
\rowcolor{LightBlue}1.559 & 0.4274 &{ 7.280} & 248.63 & 1.614 &0.06603 & 400 \\
\hline
\end{tabular}
\caption{Summary of the lattice parameters of the HotQCD $N_f=2+1$ HISQ lattices used in our study. Values of $\beta$ at which $T=0$ correlators are available are emphasized by a light blue backgorund.}  
\end{table}

Here, we report on preliminary results from our study of in-medium
charmonium using the ${\cal O} (v^4)$ NRQCD Lagrangian for the
calculation of correlator at $T \neq 0$. We use the same
$48^3 \times 12$ gauge configurations generated by HotQCD
collaboration \cite{Bazavov:2011nk,Bazavov:2014pvz} as in our bottomonium study
\cite{Kim:2014iga}). The lattice NRQCD parameters for $T \neq
0$ are listed in Table \ref{tab:latticedetail} (see
\cite{Bazavov:2011nk} for the details on how the lattice gauge fields
are generated). At $\beta = 6.664, 6.800, 6.950 $ (on $32^3 \times 32$
latices), $\beta = 6.740, 6.880, 7.030$ (on $48^3 \times 48$ lattice)
and $\beta = 7.150, 7.280$ (on $48^3 \times 64$ lattice), accompanying
$T = 0$ calculations are performed. Since $M_ca < 1$, we tested the
value $n = 6$ and 8 for the Lepage parameter and the result from $n =
6$ is used for the following analysis.

\section{Preliminary Result}

\begin{figure}[ht]
\centering
\begin{minipage}[b]{0.45\linewidth}
\includegraphics[width=\textwidth]{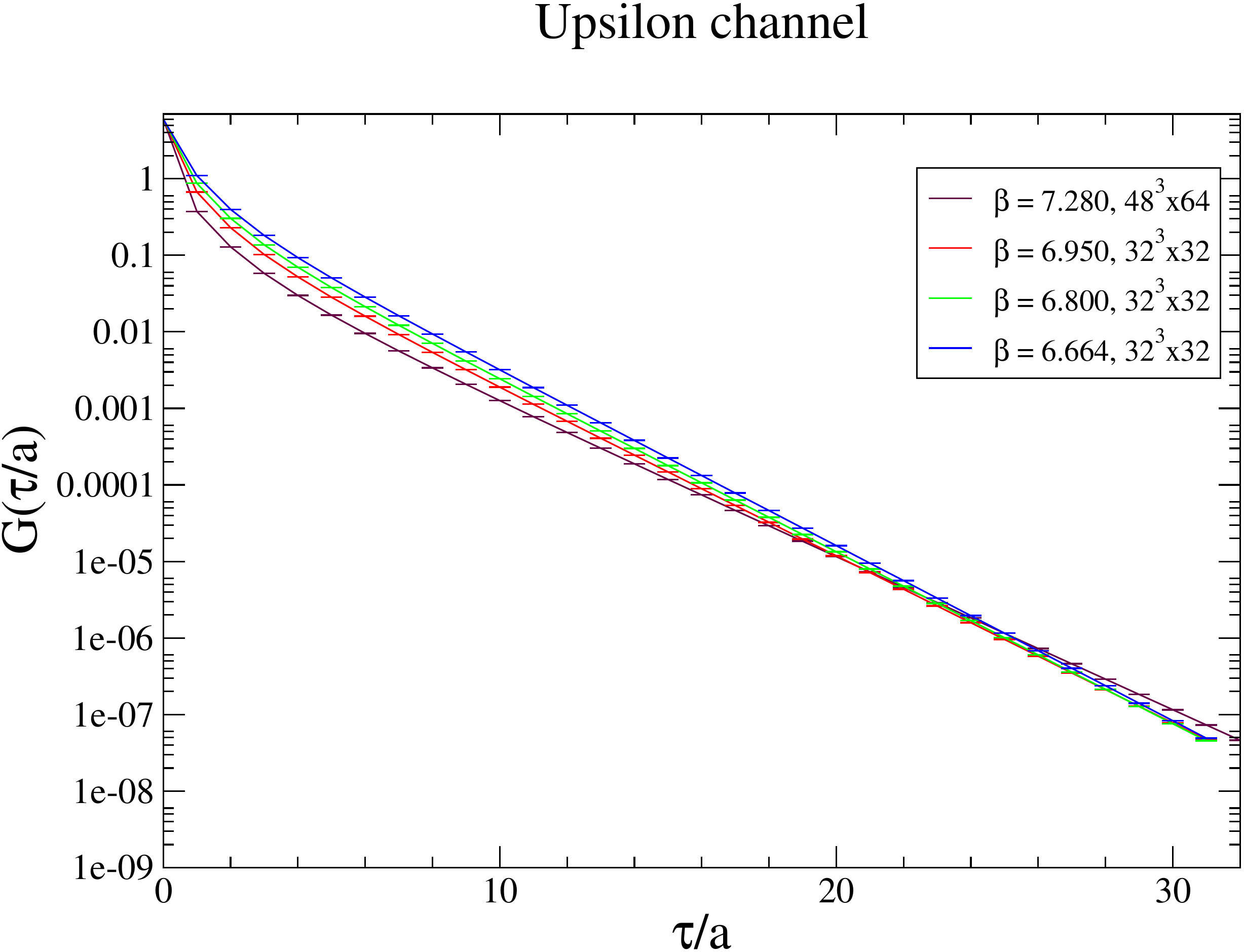}
\end{minipage}
\quad
\begin{minipage}[b]{0.45\linewidth}
\includegraphics[width=\textwidth]{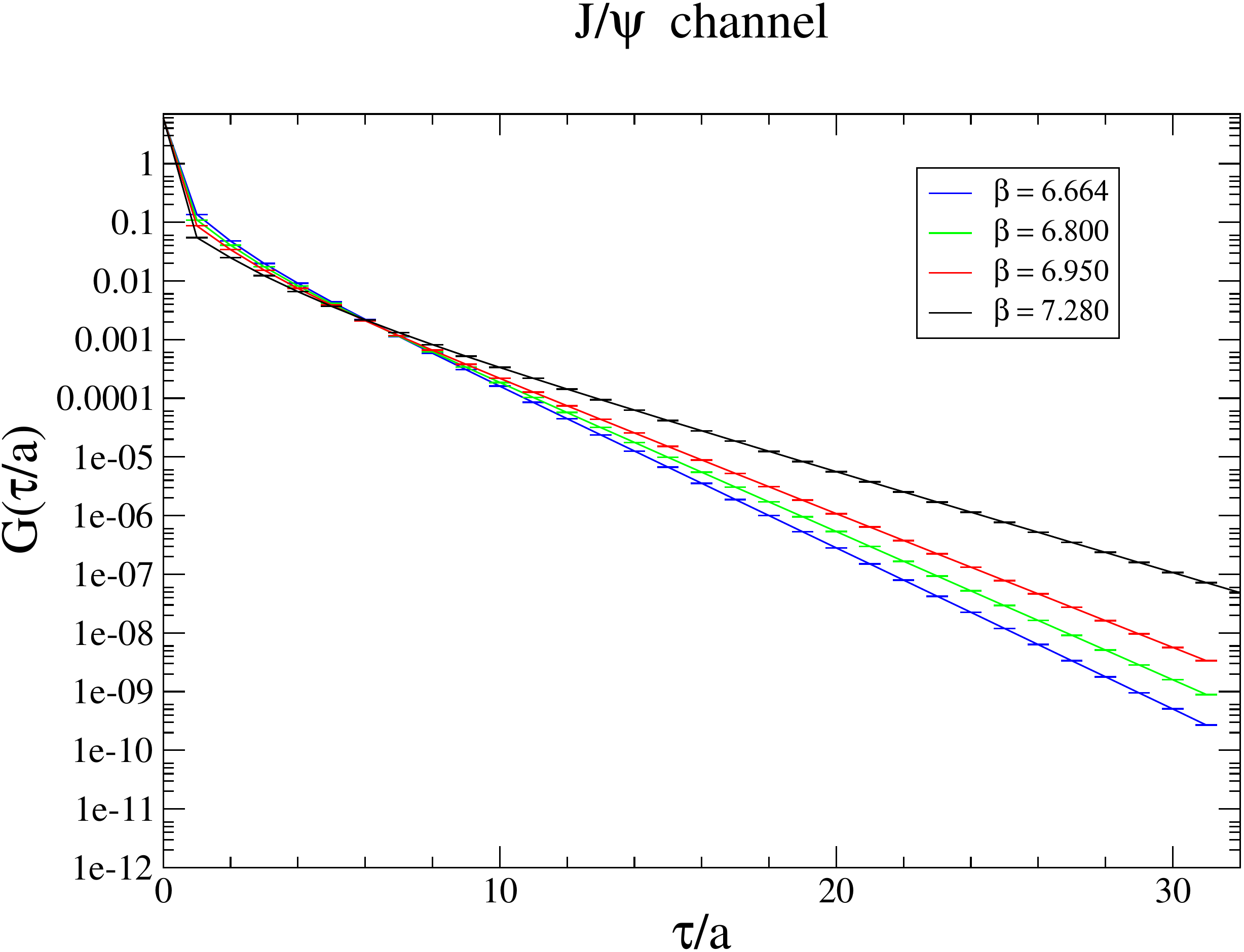}
\end{minipage}
\caption{S-wave bottomonium correlator (left) \cite{Kim:2014iga} and
  S-wave charmonium correlator (right) at $T = 0$}
\label{fig:swave}
\end{figure}

\begin{figure}[ht]
\centering
\begin{minipage}[b]{0.45\linewidth}
\includegraphics[width=\textwidth]{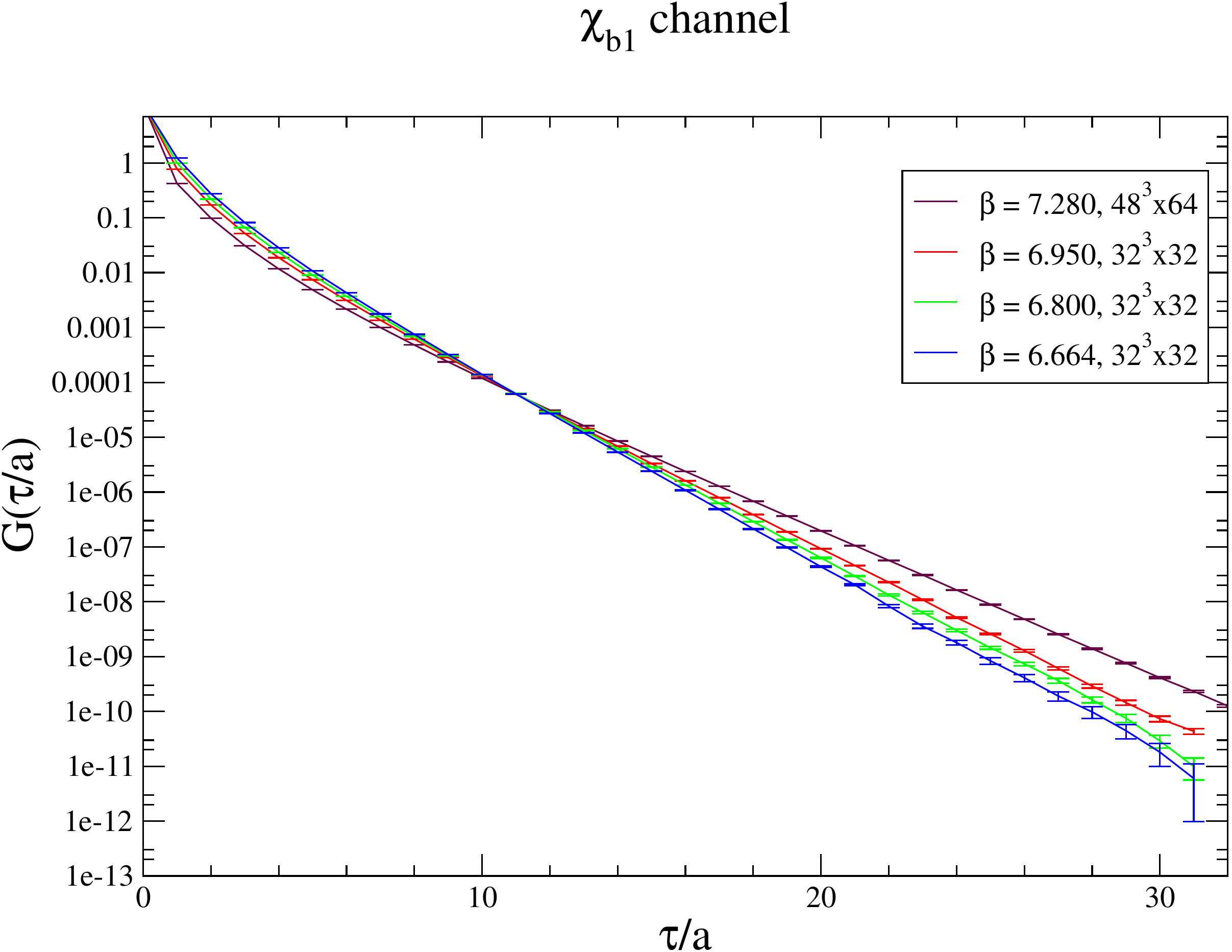}
\end{minipage}
\quad
\begin{minipage}[b]{0.45\linewidth}
\includegraphics[width=\textwidth]{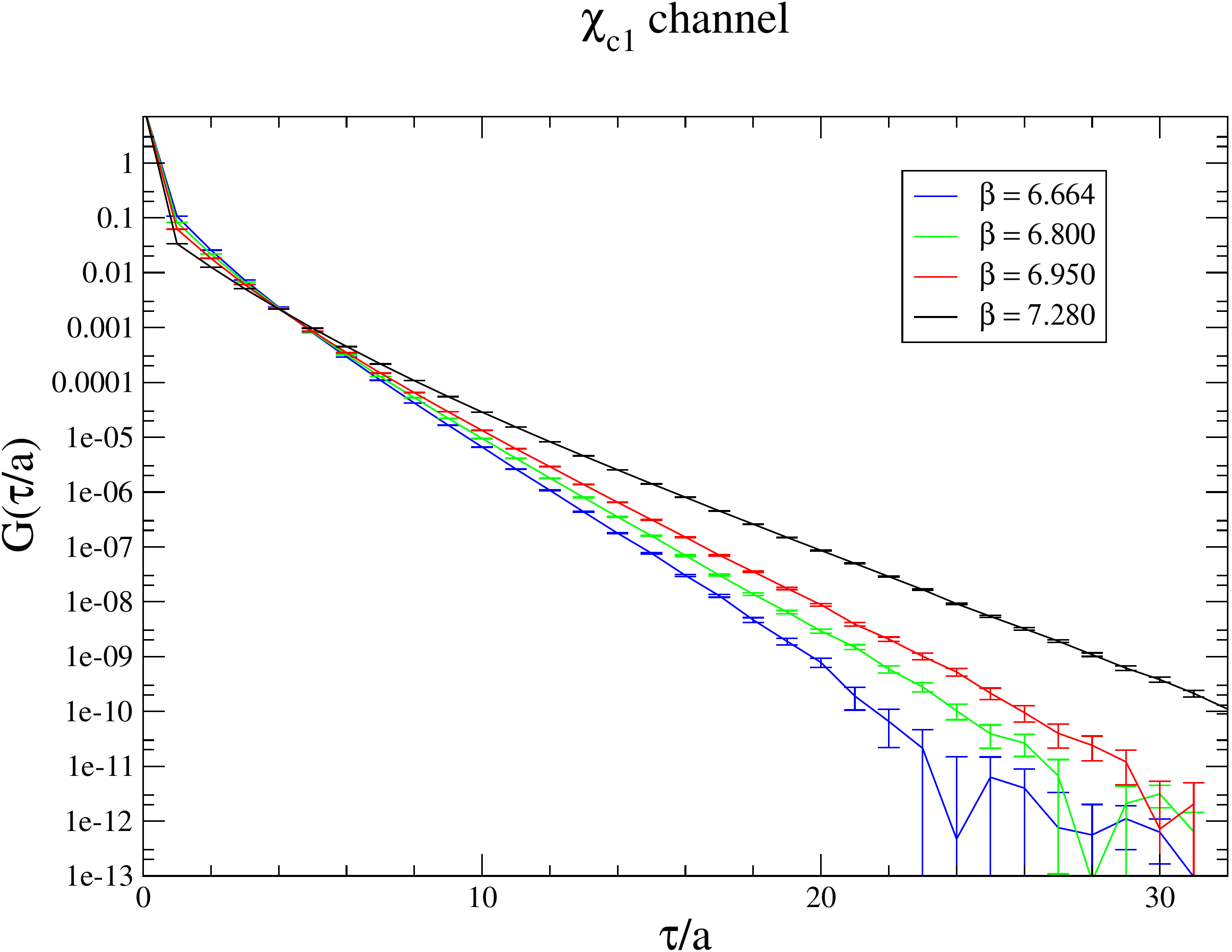}
\end{minipage}
\caption{P-wave bottomonium correlator (left) \cite{Kim:2014iga} and
  P-wave charmonium correlator (right) at $T = 0$}
\label{fig:pwave}
\end{figure}

To determine the ``zero energy shift'' in NRQCD spectra, we need $T =
0$ simulations at each $\beta$. Note that relativistic corrections are
expected to be larger in the charmonium system compared to
bottomonium, and thus the T=0 masses obtained there might differ
significantly from their PDG value.  Fig. \ref{fig:swave} shows
$J/\psi$ correlators at $T = 0$ (right) and compares them to
$\Upsilon$ correlators (left). Fig. \ref{fig:pwave} shows $\chi_{c1}$
correlators (right) and compares them to $\chi_{b1}$ correlators
(left). Both S- and P-wave charmonium NRQCD correlators at $T = 0$
follow an exponentially falling behavior and both exponential fits to
the correlators or the peak position from spectral functions
reconstructed with Bayesian methods give reasonable $\chi_{c1}-J/\psi$
splitting ($\sim 450$ MeV, compared to $(\chi_{c1}-J/\psi)_{\rm exp} \simeq
413$ MeV).

\begin{figure}[ht]
\centering
\begin{minipage}[b]{0.45\linewidth}
\includegraphics[width=\textwidth]{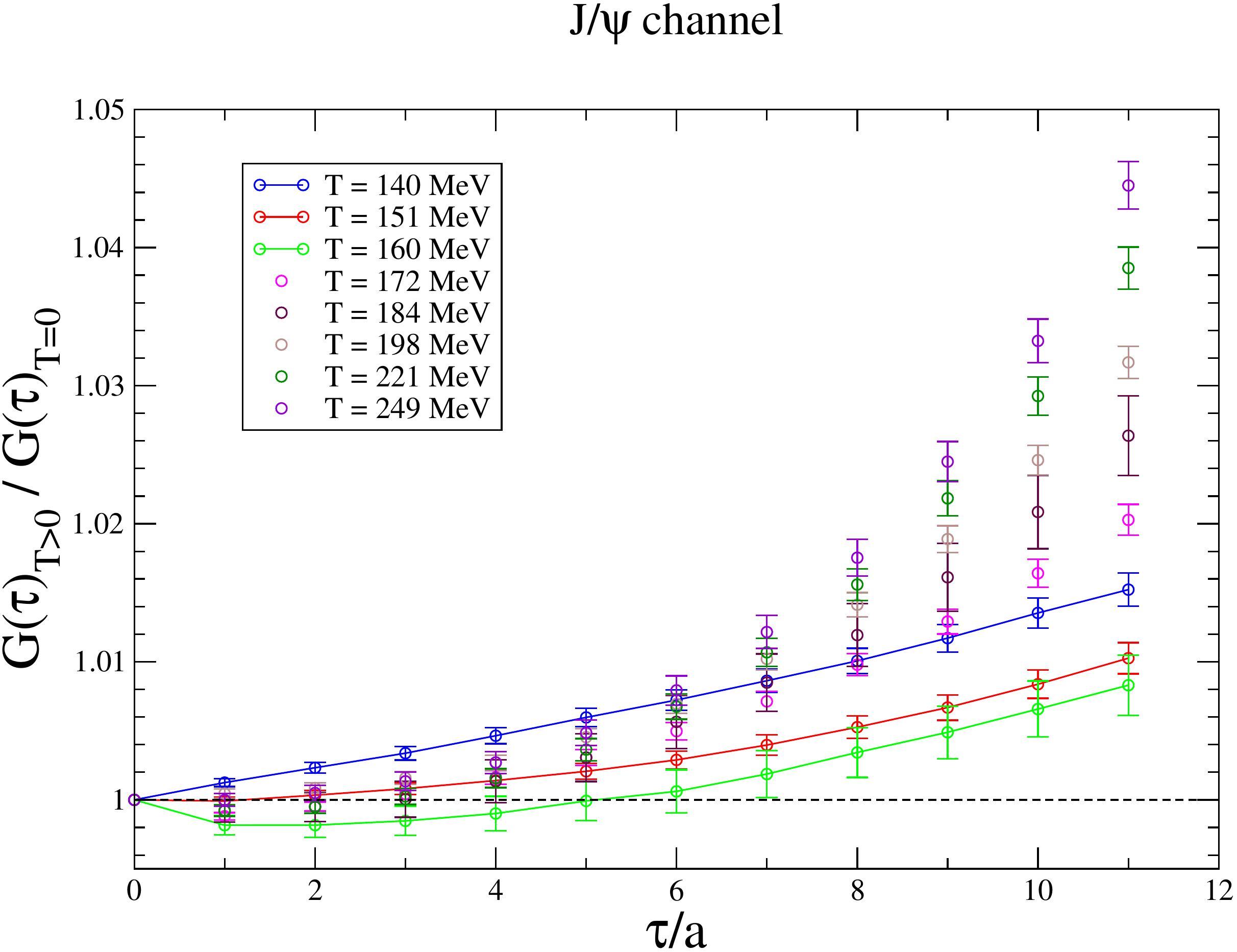}
\end{minipage}
\quad
\begin{minipage}[b]{0.45\linewidth}
\includegraphics[width=\textwidth]{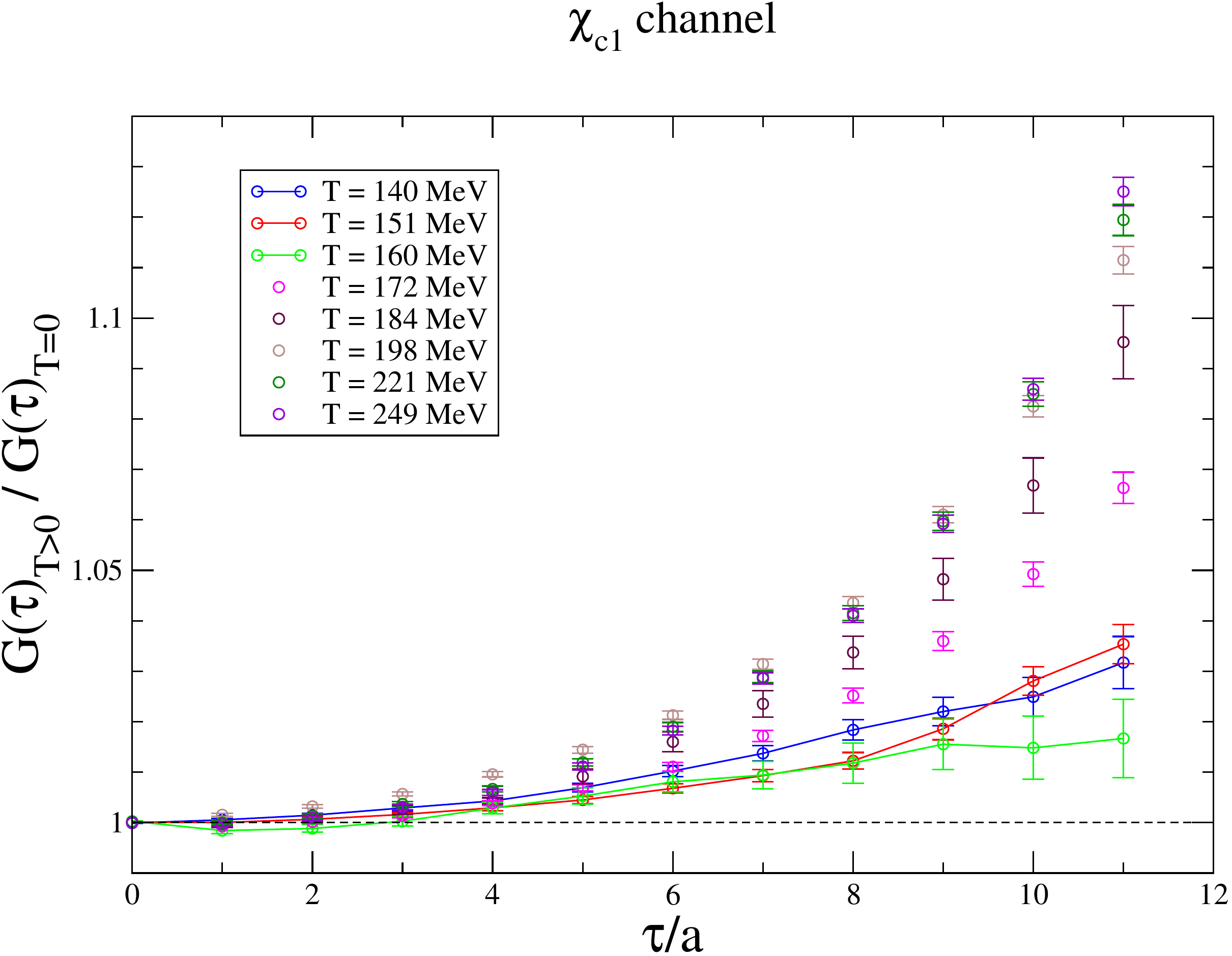}
\end{minipage}
\caption{the ratio of S-wave charmonium correlator at $T \neq 0$ to $T
  = 0$ S-wave charmonium correlator (left) and the ratio of P-wave
  charmonium correlator at $T \neq 0$ to $T = 0$ P-wave charmonium
  correlator (right). the lines connecting the data points for $T \leq
  160$ MeV are there just to guide.}
\label{fig:ratio}
\end{figure}

Fig. \ref{fig:ratio} shows the ratio of $T \neq 0$ and $T = 0$
correlators for the S- and P-wave charmonium at eight different values
of $\beta$. As expected, temperature effects in the charmonium system
are larger than those for bottomonium. Compared to bottomonium, which
shows changes up to $\sim 1 \%$ for the S-wave at the highest
temperature and up to $\sim 5 \%$ for P-wave \cite{Kim:2014iga},
charmonium shows an increase up to $\sim 5 \%$ for the S-wave at the
highest temperature and up to $\sim 13 \%$ for the
P-wave. Interestingly, both S- and P-wave charmonium correlator ratios
show non-monotonic behavior as the temperature increases. Both below
and above $T_c$ ($T_c = 159$ MeV), the ratios for S- and P-wave
charmonium are larger than one.  As the temperature approaches $T_c$
from below, the ratios decrease toward unity. As the temperature
further increases above $T_c$, the ratios also start to increase and
to move further away from the value one.  Such a non-trivial behavior
may be understood in terms of an intricate interplay of the positions
and widths of peaks in the underlying spectral functions. Lattice QCD
based potential studies \cite{Burnier:2015tda} may help to further
shed light on these phenomena.

\begin{figure}[ht]
\centering
\begin{minipage}[b]{0.45\linewidth}
\includegraphics[width=0.95\textwidth]{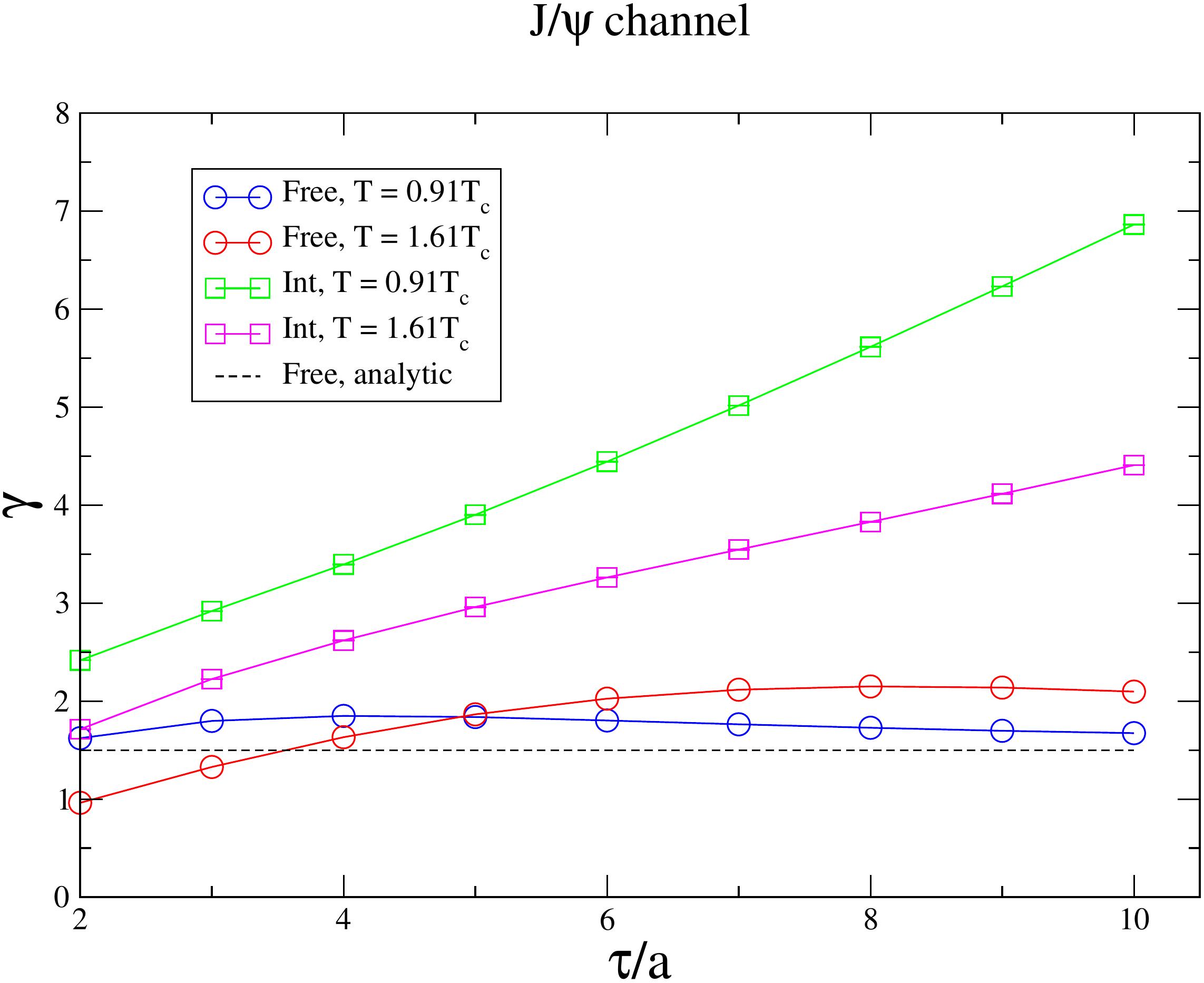}
\end{minipage}
\quad
\begin{minipage}[b]{0.45\linewidth}
\includegraphics[width=0.95\textwidth]{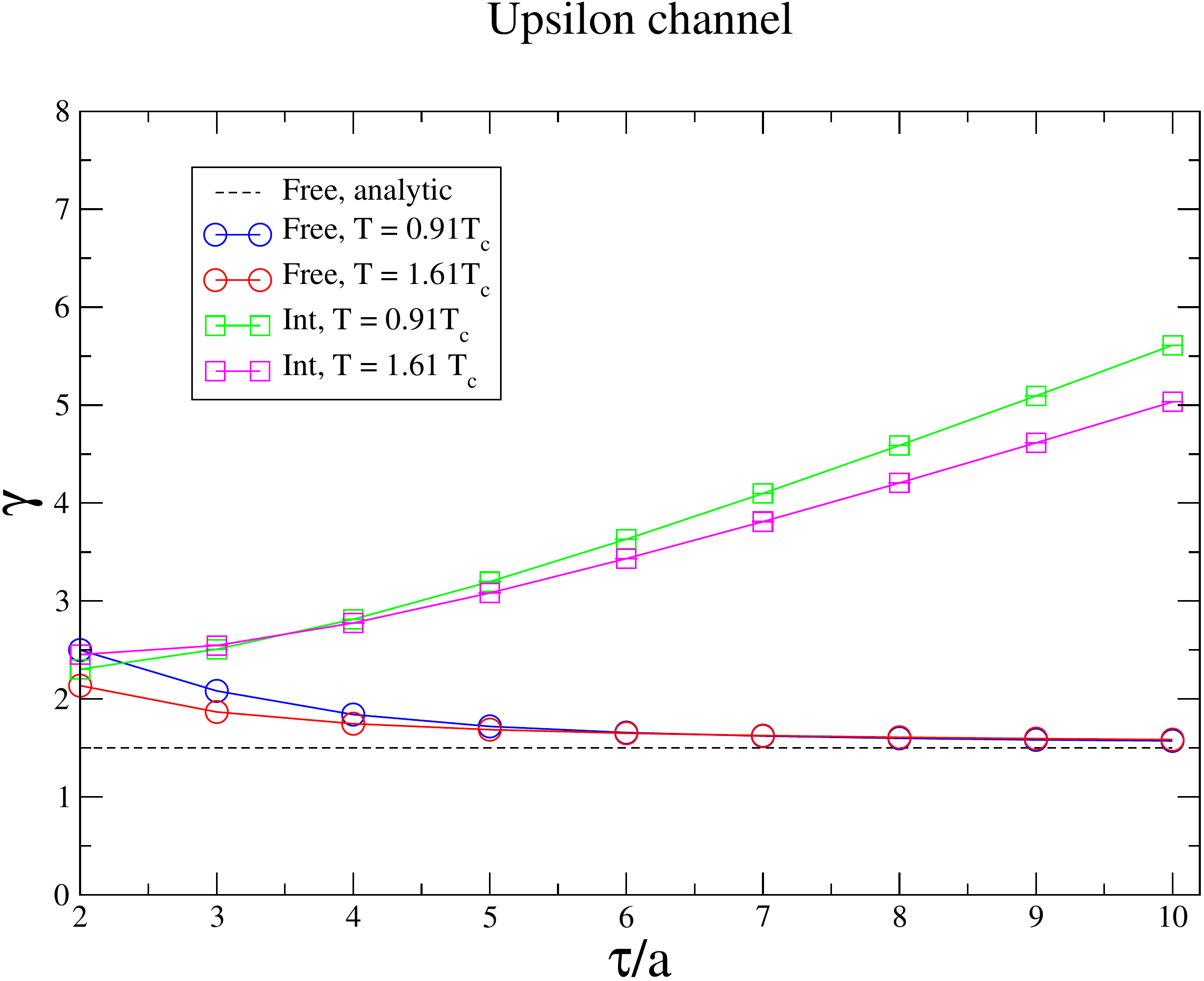}
\end{minipage}
\caption{the effective power of S-wave charmonium correlator (left) and that
  of S-wave bottomonium correlator (right)}
\label{fig:spower}
\end{figure}

\begin{figure}[ht]
\centering
\begin{minipage}[b]{0.45\linewidth}
\includegraphics[width=0.95\textwidth]{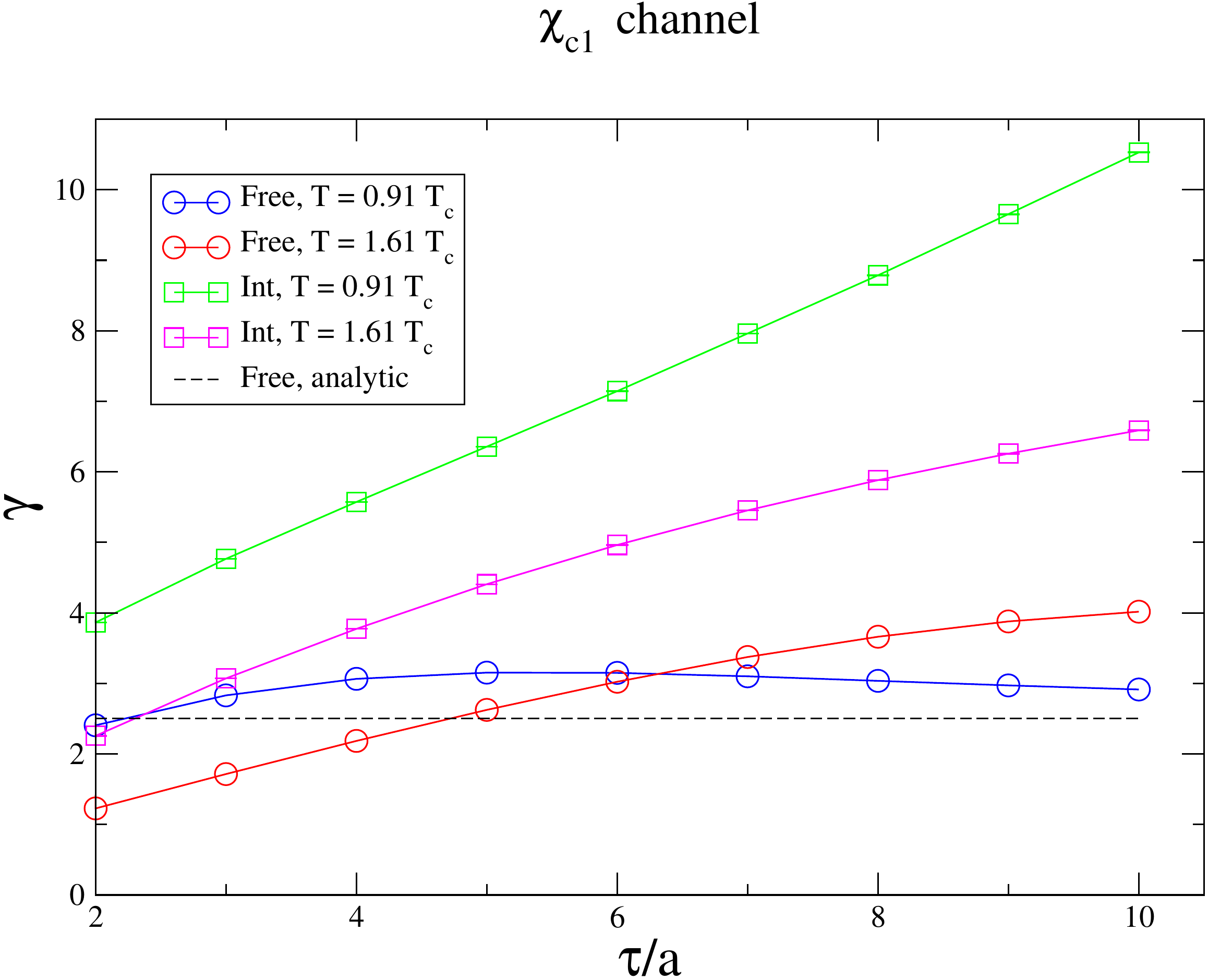}
\end{minipage}
\quad
\begin{minipage}[b]{0.45\linewidth}
\includegraphics[width=0.95\textwidth]{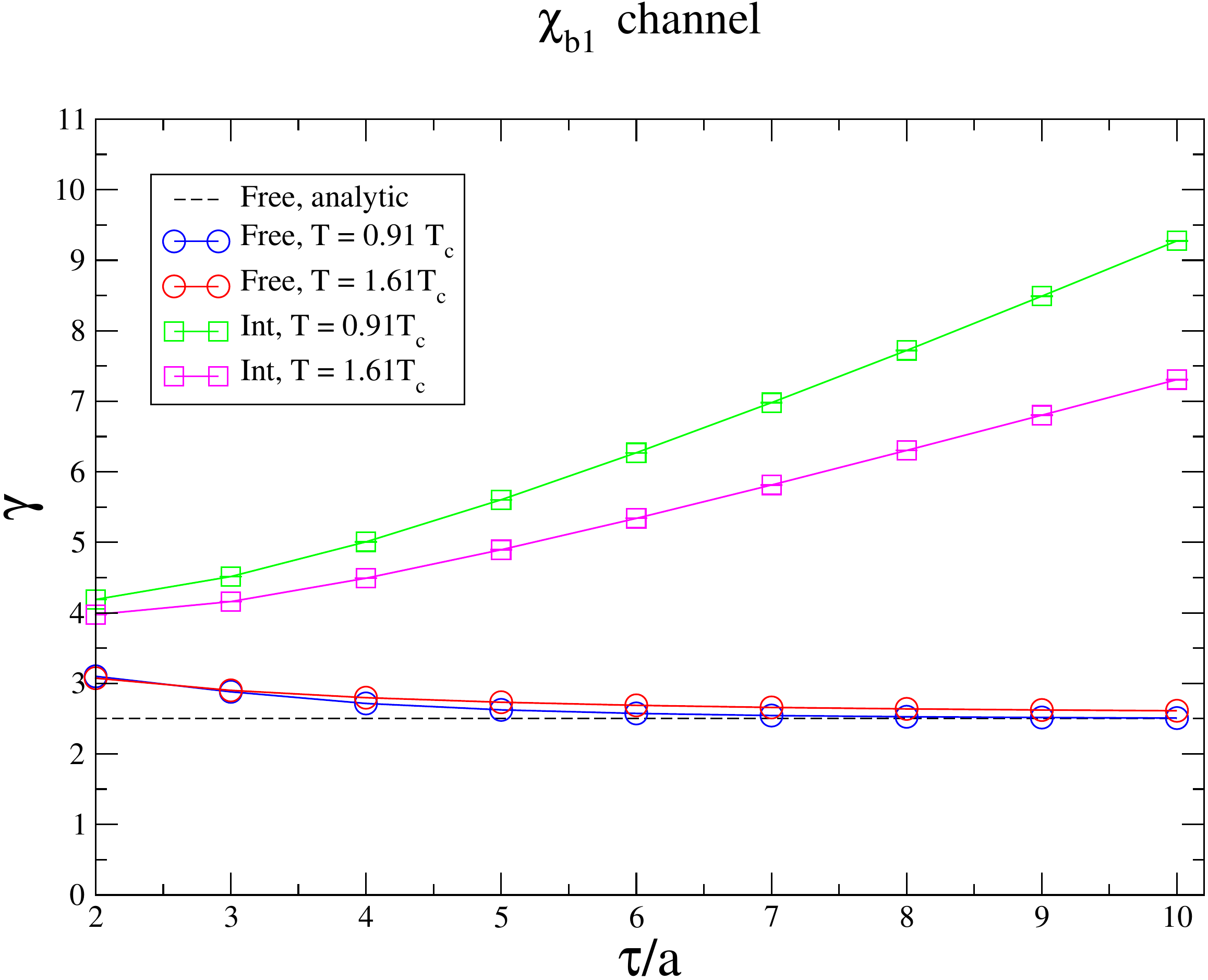}
\end{minipage}
\caption{the effective power of P-wave charmonium correlator (left) and that
  of P-wave bottomonium correlator (right)}
\label{fig:ppower}
\end{figure}

Let us turn to another measure of in-medium modification, the so-
called effective power $\gamma(\tau)$ defined from the normalized
Euclidean time derivative of the NRQCD correlators
\begin{equation}
\gamma (\tau) = \frac{\tau}{G(\tau)} \frac{dG(\tau)}{d\tau} .
\label{eq:gamma}
\end{equation}
Its values for the S-wave and P-wave channels at $T>0$ are shown in
Fig.~\ref{fig:spower} and Fig.~\ref{fig:ppower} respectively.

Each figure includes four sets of $\gamma (\tau)$. The dotted line
denotes the value for the asymptotic behavior of the
non-interacting NRQCD correlators, i.e. $3/2$ for the S-wave and $5/2$
for the P-wave \cite{Aarts:2010ek}. Two curves are calculated from the
correlators in the presence of finite T gauge fields at the lowest
($0.91 T_c$, green square) and highest temperature ($1.61 T_c$,
magenta square). Those curves belonging to non-interacting gauge
fields on trivial lattices correspond to $T = 0.91 T_c$ (i.e.,
calculated with $U_\mu (x) = 1$ but $M_c a = 0.7566$, blue circle) and
$T = 1.61 T_c$ (i.e., with $M_c a = 0.4274$, red circle). The $\gamma
(\tau)$ for free S- and P-wave bottomonium approaches the asymptotic
value of the free theory for relatively small values of $\tau$, while
$\gamma(\tau)$ for the interacting S- and P-wave bottominium are quite
different from the free behavior.

In the case of charmonium (both for S- and P-wave) the effective power
$\gamma(\tau)$ corresponding to the free lattice correlator is
different from the analytic result even for large $\tau$. These
differences are larger at high temperatures.  At $T=0.91T_c$ the
effective power is not very different for charmonium and
bottomonium. At higher temperatures, i.e at $T=1.6T_c$ there are
significant differences. We see a larger temperature dependence of
$\gamma$ for charmonium, and the value of $\gamma$ gets closer to the
free theory result. This implies stronger modification of bound state
properties and/or dissolution of the bound states in the charmonium
sector. On the other hand $\gamma$ in the interaction case is always
quite different from the free case. This is most likely due to the
non-trivial behavior of the spectral functions near the threshold
\cite{Mocsy:2007yj}. Even if all bound states are melted the free
spectral function is a poor approximation of the interacting
quarkonium spectral function \cite{Mocsy:2007yj}. Furthermore, the
interpretation of $\gamma (\tau)$ is quite subtle due to
renormalization effects at the threshold near $2 M$ and dicretization
effects in the quarkonium correlators \cite{Aarts:2014cda}. The latter
aspect is a less of a concern for our charmonium study, since the
HotQCD gauge configurations are quite close to the continuum limit.

\section{Discussion}

S- and P-wave charmonium correlators are calculated using a lattice
discretized formulation of NRQCD.  In the ratio of non-zero
temperature to $T = 0$ correlators, a larger temperature effect is
seen for charmonium system than bottomonium system. The size of
temperature effect in $J/\psi$ is similar to that in
$\chi_{b1}$. $\chi_{c1}$ shows even larger temperature effect in the
ratio.

The effective power ($\gamma$) defined by the normailized Euclidean
time derivative of the correlator suggests that the $T \neq 0$
$J/\psi$ channel is binding up to $T = 1.61 T_c (= 249)$ MeV, since
$\gamma$ of $J/\psi$, calculated on the corresponding thermal gauge
configuration behaves far differently from that of $J/\psi$ on a
non-interacting gauge configuration. $\chi_{c1}$ appears to be melting
at $T = 1.61 T_c (= 249)$ MeV, since $\gamma$ of $\chi_{c1}$
calculated on the thermal gauge configuration behaves similar to that
of $\chi_{c1}$ on a trivial gauge configuration. Of course, since a
charm quark is lighter than a bottom quark, the charmonium system may
suffer from larger relativistic corrections. $M_c a < 1$ (see
Tab. \ref{tab:latticedetail}) may cause sizeable radiative corrections
in the coefficients of the NRQCD Lagrangian. Further study is required
to understand the behavior of the effective power, in particular since
renormalization effects around the threshold can obscure its meaning
\cite{Aarts:2014cda}.

The reconstruction of spectral functions for S- and P-wave charmonium
using both MEM and the new Bayesian method, as well as the
investigation of their systematic uncertainties using increased
statistics is under way.

\vspace{0.2cm}
{\bf Acknowledgements}

SK is supported by the National Research Foundation of Korea grant
funded by the Korean government (MEST) No.\ 2012R1A1A2A04668255. PP is
supported by U.S.Department of Energy under Contract
DE-SC0012704. NRQCD correlators on HotQCD configurations are computed
using the LQCD Computing Facilities at Jefferson Lab. We thank the HotQCD
collaboration for providing the gauge configurations for this study.

\end{document}